\documentclass{emulateapj}
\shorttitle{Prompt Optical Detection of GRB~050401}
\shortauthors{Rykoff et al.}

\begin{document}

\title{Prompt Optical Detection of GRB~050401 with ROTSE-IIIa}

\author{
Rykoff,~E.~S.\altaffilmark{1}, 
Yost,~S.~A.\altaffilmark{1},
Krimm,~H.~A.\altaffilmark{2,3},
Aharonian,~F.\altaffilmark{4},
Akerlof,~C.~W.\altaffilmark{1},
Alatalo,~K.\altaffilmark{1},
Ashley,~M.~C.~B.\altaffilmark{5},
Barthelmy,~S.~D.\altaffilmark{2},
Gehrels,~N.\altaffilmark{2},
G\"{u}ver, T.\altaffilmark{6},
Horns,~D.\altaffilmark{4},
K{\i}z{\i}lo\v{g}lu,~\"{U}.\altaffilmark{7},
McKay,~T.~A.\altaffilmark{1},
\"{O}zel,~M.\altaffilmark{8},
Phillips,~A.\altaffilmark{5}, 
Quimby,~R.~M.\altaffilmark{9},
Rujopakarn,~W.\altaffilmark{1},
Schaefer,~B.~E.\altaffilmark{10}, 
Smith,~D.~A.\altaffilmark{1},
Swan,~H.~F.\altaffilmark{1},
Vestrand,~W.~T.\altaffilmark{11},
Wheeler,~J.~C.\altaffilmark{9},
Wren,~J.\altaffilmark{11}
}

\altaffiltext{1}{University of Michigan, 2477 Randall Laboratory, 450 Church
        St., Ann Arbor, MI, 48109, erykoff@umich.edu, sayost@umich.edu,
        akerlof@umich.edu, kalatalo@umich.edu, tamckay@umich.edu, wiphu@umich.edu,
        donaldas@umich.edu, hswan@umich.edu}
\altaffiltext{2}{NASA Goddard Space Flight Center, Laboratory for High Energy
        Astrophysics, Greenbelt, MD 20771, krimm@milkyway.gsfc.nasa.gov,
        scott@milkyway.gsfc.nasa.gov,gehrels@milkyway.gsfc.nasa.gov}
\altaffiltext{3}{Universities Space Research Association, 10227 Wincopin
        Circle, Suite 212, Columbia, MD 21044}
\altaffiltext{4}{Max-Planck-Institut f\"{u}r Kernphysik, Saupfercheckweg 1,
        69117 Heidelberg, Germany, Felix.Aharonian@mpi-hd.mpg.de,
        horns@mpi-hd.mpg.de}
\altaffiltext{5}{School of Physics, Department of Astrophysics and Optics,
        University of New South Wales, Sydney, NSW 2052, Australia,
        mcba@phys.unsw.edu.au, a.phillips@unsw.edu.au}
\altaffiltext{6}{Istanbul University Science Faculty, Department of Astronomy
        and Space Sciences, 34119, University-Istanbul, Turkey, 
        tolga@istanbul.edu.tr}
\altaffiltext{7}{Middle East Technical University, 06531 Ankara, Turkey,
        umk@astroa.physics.metu.edu.tr}
\altaffiltext{8}{\c{C}anakkale Onsekiz Mart \"{U}niversitesi, Terzio\v{g}lu
        17020, \c{C}anakkale, Turkey, m.e.ozel@comu.edu.tr}
\altaffiltext{9}{Department of Astronomy, University of Texas, Austin, TX
        78712, quimby@astro.as.utexas.edu, wheel@astro.as.utexas.edu}
\altaffiltext{10}{Department of Physics and Astronomy, Louisiana State
        University, Baton Rouge, LA 70803, schaefer@lsu.edu}
\altaffiltext{11}{Los Alamos National Laboratory, NIS-2 MS D436, Los Alamos, NM
        87545, vestrand@lanl.gov, jwren@nis.lanl.gov}

\begin{abstract}
The ROTSE-IIIa telescope at Siding Spring Observatory, Australia, detected
prompt optical emission from \emph{Swift} GRB~050401.  In this letter, we
present observations of the early optical afterglow, first detected by the
ROTSE-IIIa telescope 33~s after the start of $\gamma$-ray emission,
contemporaneous with the brightest peak of this emission.  This GRB was neither
exceptionally long nor bright.  This is the first prompt optical detection of a
GRB of typical duration and luminosity.  We find that the early afterglow decay
does not deviate significantly from the power-law decay observable at later
times, and is uncorrelated with the prompt $\gamma$-ray emission.  We compare
this detection with the other two GRBs with prompt observations, GRB~990123 and
GRB~041219a.  All three bursts exhibit quite different behavior at early times.

\end{abstract}

\section{Introduction}

The detection of prompt optical emission contemporaneous with gamma-ray bursts
(GRBs) has been quite difficult.  Until now, only two bursts, GRB~990123 and
GRB~041219a, have had optical light detected while detectable $\gamma$-rays
were still being emitted.  The ROTSE-I instrument detected a bright $9^{th}$
magnitude flash coincident with GRB~990123, a burst exceptionally luminous in
$\gamma$-rays~\citep{abbbb99}.  The RAPTOR-S telescope detected faint optical
emission from GRB~041219a that was correlated with the $\gamma$-ray
emission~\citep[henceforth, V05]{vwwfs05}. GRB~041219a was an unusually long
burst (over 6 minutes) that allowed extended optical monitoring during the
$\gamma$-ray emission.  The \emph{Swift} detection of GRB~050401 and rapid
dissemination of its coordinates enabled the first prompt detection of an
optical counterpart for a GRB with a typical duration and fluence.  With a T90
of $33\,\mathrm{s}$ and a fluence of
$1.4\times10^{-5}\,\mathrm{erg}\,\mathrm{cm}^{-2}$ in the 15-350 keV
band~\citep{sbbcf05}, this burst was neither especially long nor bright.

In this letter, we report on the prompt detection of the optical afterglow of
GRB~050401 with the ROTSE-IIIa (Robotic Optical Transient Search Experiment)
telescope located at Siding Spring Observatory (SSO), Australia.  Our initial
detection of the afterglow is coincident with the brightest peak in the
$\gamma$-ray emission.  ROTSE-IIIa followed the afterglow through the first
four minutes after the burst, recording a fading afterglow consistent with a
backward extrapolation of the afterglow measured at much later times.  We
compare these observations to the two previously observed cases of prompt
optical emission, and to the empirical model of V05 that suggested a coupling
of $\gamma$-ray and optical flux.

\section{Observations and Analysis}
\label{sec:observations}

The ROTSE-III array is a worldwide network of 0.45~m robotic, automated
telescopes, built for fast ($\sim 6$ s) responses to GRB triggers from
satellites such as HETE-2 and \emph{Swift}.  They have wide ($1\fdg85 \times
1\fdg85$) fields of view imaged onto Marconi $2048\times2048$ back-illuminated
thinned CCDs, and operate without filters.  The ROTSE-III systems are described
in detail in \citet{akmrs03}.

On 2005 April 01, \emph{Swift}/BAT detected GRB~050401 (\emph{Swift} trigger
113120) at 14:20:15 UT.  The position was distributed as a Gamma-ray Burst
Coordinates Network (GCN) notice at 14:20:34 UT, with a $4\arcmin$ radius error
box.  The burst had a $T_{90}$ duration of $33\pm2\,\mathrm{s}$, and the
position was released during the $\gamma$-ray emission~\citep{sbbcf05}. The
\emph{Swift} trigger time was 9 seconds after the start of the GRB; in this
paper we reference all times to the start of $\gamma$-ray emission at 14:20:06
UT.

ROTSE-IIIa responded automatically to the GCN notice, beginning its first
exposure in less than 6 seconds, at 14:20:39.2 UT, during the largest peak of
the $\gamma$-ray emission.  The automated burst response included a set of ten
5-s exposures, ten 20-s exposures, and a long sequence of 60-s exposures
continuing for about 5 hours until twilight.  Initial analysis of the prompt
response did not yield an obvious afterglow candidate.  About an hour after the
burst, at 15:17:16.8 UT, \citet{mp05} initiated a burst response on the SSO
40-inch telescope.  They detected a new $20^{th}$ magnitude object at
$\alpha=16^h31^m28\fs8$, $\delta=02\arcdeg11\arcmin14\farcs2$ (J2000.0) which
they identified as the optical counterpart.  Further analysis of the ROTSE-IIIa
images revealed this source at magnitudes close to our detection limit.  Later
spectroscopic observations by \citet{fjhws05} at the VLT revealed a redshift of
2.9 for this burst.  The burst position has a high galactic latitude of
$31\fdg8$, so extinction from the Milky Way is not significant.

\begin{figure}
\rotatebox{90}{\scalebox{0.85}{\plotone{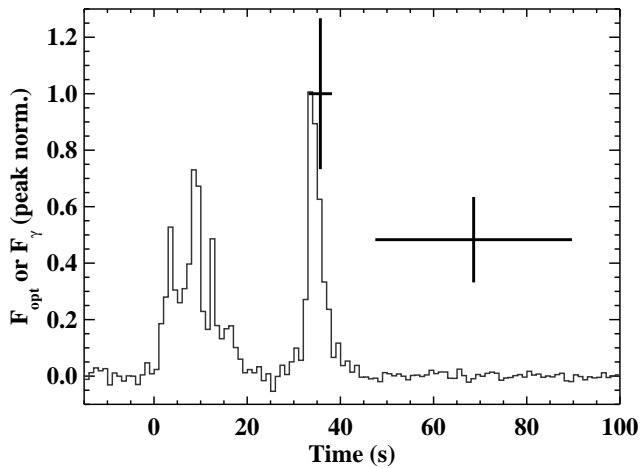}}}
\caption{\label{fig:grblc}$\gamma$-ray lightcurve for GRB~050401.  The time is
  seconds since the start of $\gamma$-ray emission at 14:20:06 UT.  The burst
  T90 duration was $33\pm2\,\mathrm{s}$.  The first two optical detections
  (peak normalized) have been overplotted.  The first ROTSE-IIIa observation is
  coincident with the brightest $\gamma$-ray peak, and there is no correlation
  between the $\gamma$-ray flux and the optical flux at the early time.}
\end{figure}

The $\gamma$-ray light curve from the \emph{Swift}/BAT instrument is shown in
Figure~\ref{fig:grblc}.  The light curve has been normalized to the peak flux.
Overplotted are the first two ROTSE-IIIa observations, with the first 5-s
integration coincident with the brightest peak in the $\gamma$-ray emission.
That burst was 56 degrees from the spacecraft axis, which means that the source
illuminated only 8\% of the BAT detectors~\citep{bbcfg05}.  The \emph{Swift}
spacecraft began its slew to the target during the ROTSE-IIIa observation,
delayed by 9 seconds due to an earth-limb constraint.  All the BAT flux values
were corrected for partial illumination and other geometric effects, including
the spacecraft slew. The $\gamma$-ray spectrum during this period is well fit
by a simple power law with a photon index of $1.58\pm0.06$, with a $\chi^2$ of
58.0 (57 d.o.f.).  This is consistent with the index early in the burst
suggesting that there is no significant spectral
evolution. Table~\ref{tab:fluxvals} shows the flux density and flux
measurements for the $\gamma$-ray emission coincident with the first two
ROTSE-IIIa observations. To obtain a $3\sigma$ upper limit for the $\gamma$-ray
flux coincident with the second ROTSE-IIIa integration, we assumed the source
had the same spectral shape as the first integration.

\begin{deluxetable}{cccc}
\tablewidth{0pt}
\tablecaption{Simultaneous ROTSE-III and \emph{Swift} measurement of
  GRB~050401.\label{tab:fluxvals}}
\tabletypesize{\scriptsize}
\tablehead{
\colhead{Obs.} &
\colhead{Energy Band} & \colhead{Flux Density (mJy)} & \colhead{Flux
  ($\mathrm{erg}\,\mathrm{cm}^{-2}\,\mathrm{s}^{-1}$)}
}
\startdata
1 & $R_c$-band\tablenotemark{a} & $0.59\pm0.16$ & $6.6\pm1.8\times10^{-13}$\\ 
& 15-350 keV &              & $7.60\pm24\times10^{-7}$\\
& 15-25 keV & $2.73\pm0.09$ & $6.60\pm0.21\times10^{-8}$ \\
& 25-50 keV & $1.91\pm0.06$ & $1.16\pm0.04\times10^{-7}$ \\
& 50-100 keV & $1.28\pm0.12$ & $1.55\pm0.14\times10^{-7}$ \\
& 100-350 keV & $0.70\pm0.02$ & $4.24\pm0.14\times10^{-7}$\\
\hline
2 & $R_c$-band\tablenotemark{a} & $0.28\pm0.08$ & $3.2\pm0.9\times10^{-13}$\\
& 15-350 keV &               & $<4.02\times10^{-8}$\\
\enddata
\tablenotetext{a}{The unfiltered ROTSE magnitudes have been calibrated such
  that they are roughly equivalent to the $R_c$ band system.}
\tablecomments{Observation 1 is 33.2~s - 38.2~s post-burst, and observation 2
  is 47.5~s - 89.7~s post-burst.}
\end{deluxetable}

The ROTSE-IIIa images were bias-subtracted and flat-fielded.  The flat-field
image was generated from 30 twilight images.  We used SExtractor~\citep{ba96}
to perform the initial object detection and to determine the centroid positions
of the stars.  After the first 5-s integration, images were co-added in sets of
three to improve our signal to noise.  The transient is not detected in
individual frames, which have limits consistent with the magnitudes derived
from the co-added frames.  The images were then processed with a customized
version of the DAOPHOT PSF fitting package~\citep{stetson87} that has been
ported to the IDL Astronomy User's Library~\citep{landsman95}.  The magnitude
zero-point for each image is calculated from the median offset to the USNO~1-m
$R$-band standard stars~\citep{henden05} in the magnitude range of
$13.5<V<20.0$ with $0.4 < V-R < 1.0$.  As we have no data on afterglow color
information at the early time, no additional color corrections have been
applied to our unfiltered data.

\begin{figure}
\rotatebox{90}{\scalebox{0.60}{\plotone{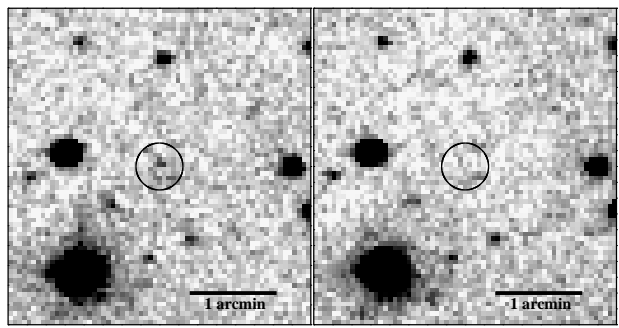}}}
\caption{\label{fig:mosaic}Optical counterpart of GRB~050401.  The panel on the
  left shows the counterpart in a co-added image from 33~s to 281~s
  post-burst.  The counterpart is absent from the panel on the right, a
  co-added image from 290~s to 487~s post-burst.}
\end{figure}

Figure~\ref{fig:mosaic} shows the optical counterpart and a later non-detection
image. The panel on the left is a co-addition of all our images with
significant flux, from 33~s to 281~s post-burst.  The panel on the right is the
subsequent non-detection image from 290~s to 487~s
post-burst. Table~\ref{tab:photometry} contains the optical photometry for the
early afterglow.  In addition, Table~\ref{tab:fluxvals} shows the approximate
flux density for our first two observations, assuming the ROTSE-IIIa unfiltered
magnitudes are roughly equivalent to the $R_c$-band system.

\begin{deluxetable}{lcrrc}
\tablewidth{0pt}
\tablecaption{Optical Photometry for GRB~050401\label{tab:photometry}}
\tabletypesize{\scriptsize}
\tablehead{
  \colhead{Telescope} &
  \colhead{Filter} &
  \colhead{$t_{\mathrm{start}}$ (s)} &
  \colhead{$t_{\mathrm{end}}$ (s)} &
  \colhead{Magnitude}
}
\startdata
ROTSE-IIIa & None &        33.2 &         38.2 & $16.80\pm 0.29$\\
ROTSE-IIIa & ... &         47.5 &         89.7 & $17.59\pm 0.34$\\
ROTSE-IIIa & ... &         99.2 &        140.9 & $17.42\pm 0.23$\\
ROTSE-IIIa & ... &        150.2 &        184.3 & $17.88\pm 0.25$\\
ROTSE-IIIa & ... &        201.5 &        281.2 & $18.58\pm 0.43$\\
ROTSE-IIIa & ... &        290.3 &        487.1 & $>18.60$\\
\enddata
\tablecomments{All times are in seconds since the burst time, 14:20:06 UT (see \S~\ref{sec:observations})}
\end{deluxetable}

\section{Results}

\begin{figure}
\rotatebox{90}{\scalebox{0.85}{\plotone{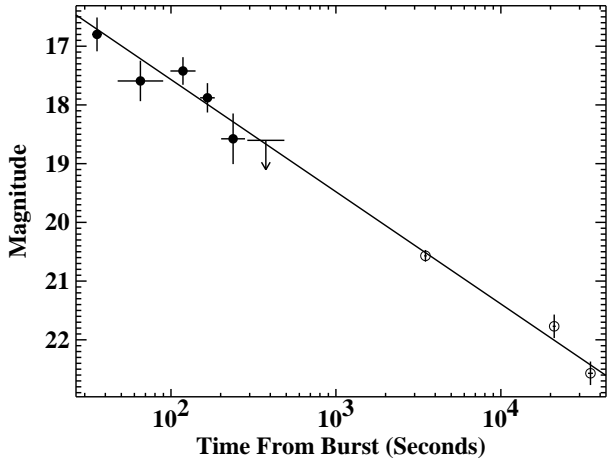}}}
\caption{\label{fig:lightcurve}Optical lightcurve for GRB~050401.  The filled
  circles are from ROTSE-IIIa and the empty circles are taken from the
  literature~\citep{pm05,kispr05,mkp05}.  A power-law fit with a decay
  slope of $\alpha=-0.76\pm0.03$ is overplotted.  The early
  optical afterglow, including the first point coincident with the $\gamma$-ray
  emission, does not show any significant deviation from the power-law decay
  visible at later times.}
\end{figure}

Figure~\ref{fig:lightcurve} shows the optical light curve of GRB~050401 with
the ROTSE-IIIa observations combined with later followup from larger
telescopes.  The light curve for the first 40000~s is well fit by a single
power-law $f_\nu\propto{t^{\alpha}}$ with a decay slope $\alpha=-0.76\pm0.03$
($\chi^2=4.7$, 6 d.o.f.).  Interestingly, there is no evidence that the
afterglow is either brighter or dimmer during the prompt $\gamma$-ray emission
than one would predict from an extrapolation of the later afterglow.  We see no
evidence for excess emission expected from a reverse shock flash, nor do we see
evidence for a deficit of emission during the rise of the early afterglow.

With the detection of a prompt optical counterpart, we can compare the optical
to $\gamma$-ray flux ratio from GRB~050401 to that of GRB~041219a and
GRB~990123, the other two bursts with prompt detections.  Following the method
of V05, we have calculated the optical to $\gamma$-ray flux ratio
$F_{R_c}/F_\gamma$ for the prompt optical observation and the first subsequent
co-added integration. As with V05, we use the flux integrated in the
\emph{Swift}/BAT 15-350 keV band over the duration of our observation. We have
not performed any $k$-corrections because we do not know the spectral shape of
the prompt optical emission.

The flux ratio for the first ROTSE-IIIa observation of GRB~050401 is
$8.7\pm2.3\times10^{-7}$.  We have tested the correlation between the optical
and $\gamma$-ray fluxes for the first two integrations.  We fit a simple
proportional model of the form $F_\mathrm{opt} = aF_\gamma$ using the two
optical detections and the $\gamma$-ray detection and upper limit.  This
proportional model results in a very poor fit, with a $\chi^2$ probability of
0.04\%.  Therefore, the $\gamma$-ray and optical flux are not correlated.


The flux ratio during the first ROTSE-III integration is $\sim$14 times dimmer
in optical than that calculated for GRB~041219a in V05.  If the flux ratio for
GRB~050401 were the same as that for GRB~041219a, we would expect an optical
detection at $\sim14$ magnitude.  If the transient had been this bright we
would have detected it with a S/N of over 25, which we can firmly rule out.  In
addition, V05 had to perform an approximate galactic reddening correction of
4.9 magnitudes, and suggest that the true extinction value may be larger.  This
would imply that the optical to $\gamma$-ray flux is even larger for
GRB~041219a, and V05 would predict a brighter counterpart for GRB~050401.

We have also compared the prompt optical flux from GRB~050401 to that from
GRB~990123.  Although the optical emission from GRB~990123 is not correlated
with the $\gamma$-ray emission, V05 have suggested that the first detection of
the transient at $11^{th}$ magnitude might be related to the brightest
$\gamma$-ray peak.  Using the GRB (``Band'') model parameters from
\citet{bbkpk99}, we have calculated the flux in the 15-350 keV band for the
first ROTSE-I integration of GRB~990123.  We find that the
optical-to-$\gamma$-ray flux ratio is $1.7\times10^{-5}$, or about a factor of
20 larger than that for GRB~050401.  However, it is reasonable to expect that
this first optical detection of GRB~990123 is the onset of the reverse shock,
which is not evident in the early afterglow of GRB~050401.


The primary difficulty in comparing the optical flux to the $\gamma$-ray flux
is that all three bursts have different spectral shapes in the $\gamma$-ray
regime.  Comparing the optical and $\gamma$-ray flux densities avoids the
integration over the arbitrary $\gamma$-ray passband and can simplify the
comparison of these different bursts.  Table~\ref{tab:fluxdens} shows the flux
density at 1.9 eV (the peak of the $R_c$ passband), 20 keV, and 100 keV for the
three bursts.  We have chosen to examine the first optical integration of
GRB~990123, which might be before the onset of the reverse shock; the third
optical integration of GRB~041219a, which is coincident with the final peak in
the $\gamma$-ray emission; and the first optical integration of GRB~050401,
also coincident with the final $\gamma$-ray peak.  There does not seem to be
any obvious pattern common to all three bursts.

\begin{deluxetable*}{ccccc}
\tablewidth{0pt}
\tablecaption{Flux densities for prompt counterparts.\label{tab:fluxdens}}
\tabletypesize{\scriptsize}
\tablehead{
\colhead{Burst} & \colhead{$F_{\mathrm{opt}}
  (\mathrm{erg}\,\mathrm{cm}^{-2}\,\mathrm{s}^{-1})$} &
\colhead{$f_\nu$ [1.9 eV] (mJy)} &
\colhead{$f_\nu$ [20 keV] (mJy)} & \colhead{$f_\nu$ [100 keV] (mJy)}
}
\startdata
GRB~990123 (1) & $1.0\pm0.1\times10^{-10}$ & $89\pm12$ & $3.4\pm0.3$ & $5.7\pm0.3$ \\
GRB~041219a (3) & $4.3\pm0.9\times10^{-12}$ & $3.8\pm0.8$ & $2.88\pm0.07$ & $0.83\pm0.04$\\ 
GRB~050401 & $6.6\pm1.8\times10^{-13}$ & $0.59\pm0.16$ & $2.73\pm0.09$ & $0.99\pm0.12$\\
\enddata
\end{deluxetable*}

\section{Discussion}

Although V05 have seen evidence for correlation between the optical flux and
$\gamma$-ray flux for GRB~041219a, this correlation is absent in GRB~050401.
Each of the three GRBs with prompt optical detections displays a different
relationship between optical and $\gamma$-ray flux.  For GRB~990123, the
optical and $\gamma$-ray emission vary independently, and the optical emission
is much brighter than a back extrapolation of the afterglow would suggest.  For
GRB~041219a, the optical and $\gamma$-ray emission are correlated, but we do
not have any further observations to compare this to the later afterglow.
Finally, for GRB~050401, the optical and $\gamma$-ray emission vary
independently, and the prompt optical emission is well fit by a backward
extrapolation of the later afterglow emission.

As the prompt optical emission of GRB~050401 is indistinguishable from the
later afterglow, it is most likely radiated from the same emitting region.  In
the fireball model~\citep{p05}, the afterglow radiation is from the forward
external shock.  This would indicate that any optical emission related to the
prompt $\gamma$-ray emission radiated by the internal shocks is negligible
compared to the forward shock emission.  As the optical observations began only
33~s after the start of the $\gamma$-ray emission, this would imply a very
rapid rise in the forward shock emission.  Therefore, the typical synchrotron
peak, $\nu_m$, must have passed the optical band at $<30~\mathrm{s}$.  This is
consistent with both an ISM environment~\citep{se01} and a wind
environment~\citep{cl00} with small but reasonable values for the microphysical
parameters.  In addition, the lack of a reverse shock signature is consistent
with a high density wind medium~\citep{np04}.  This early behavior is quite
different from the behavior of GRB~990123, GRB~041219a, and for other early
afterglows such as that from GRB~030418~\citep{rspaa04} that have been observed
to rise after tens or hundreds of seconds.

The rapid localization of GRB~050401 by \emph{Swift}, combined with the rapid
response of the ROTSE-III instruments, has allowed, for the first time, the
detection of a prompt optical counterpart of a typical GRB.  \emph{Swift} will
localize $\sim75$ bursts per year, and the ROTSE-III instruments can promptly
respond to $\sim40\%$ of these bursts.  Many of these localizations will be
during the $\gamma$-ray emission, and we expect the ROTSE-III instruments to
achieve $\sim5$ prompt detections per year.  During the next few years we will
sample the range of prompt optical emission from GRBs, perhaps revealing
patterns which will inform our understanding of the underlying GRB engine.

\acknowledgements

This work has been supported by NASA grants NNG-04WC41G and NGT5-135, NSF
grants AST-0407061, the Australian Research Council, the University of New
South Wales, and the University of Michigan.  Work performed at LANL is
supported through internal LDRD funding.  Special thanks to the observatory
staff at Siding Spring Observatory.

\newcommand{\noopsort}[1]{} \newcommand{\printfirst}[2]{#1}
  \newcommand{\singleletter}[1]{#1} \newcommand{\switchargs}[2]{#2#1}

\end{document}